# The carrying of the surface electrons in quasi-one-dimensional system over superfluid helium in conditions of the negatively charged substrate.

*Nikolaenko V.A. and Smorodin A.V.*

B.Verkin Institute for Low Temperature Physics and Engineering of the National Academy of Sciences of Ukraine

**ABSTRACT**

The quasi-one-dimensional system of surface electrons over superfluid helium with negative charge on substrate is investigated. The temperature dependence of conductivity is ladder-like which is observed at temperature lower 1.3 K. The parameters of ladder have been different in different experiments. The observed dependence isn't explained by classical models of carry with account to the usual scatters in channel. It is supposed the dependence have been connected with quantum character of electron transport in narrow channels.

### Introduction

The development of modern nanotechnology stimulates the fundamental and applied investigations of one-dimensional and zero-dimensional systems. These systems known as a quantum wire and quantum dot (QD) are broadly used in solid state nanoelectronic for building SET (single electron transistor), laser on QDs etc. In the perspective there is the plan to use QD as quantum bit (QB) for quantum computer (QC) [1].

The classical analog to solid state nanostructures is structure with surface electrons (SE) [2]. The electrons over smooth dielectric substrate, with small permittivity form the two dimensional system of SE. Small potential dimple lead to mezoscopic distance of SE layer above substrate.

The superfluid helium is an ideal substrate for creating SE. The electron spectrum in normal direction is quantized as $\theta^2 R / n^2$, with energy of a basic state about 8 K (here $n$ = 1; 2; …; $R$ = 13,6 eV is Ridberg's constant; $\theta = \frac{\varepsilon-1}{4(\varepsilon+1)}$ is the effective force of the imaginary in substrate, $\varepsilon$ is dielectric constant of $^4$He and equal 1,056 ). The spectrum is a quasi-continuum along surface and the electron mobility ( ~ $10^7$ cm$^2$ / V·c) is limited only by interaction of the electrons with helium atoms in gas and ripplons (Relay's surface waves on superfluid helium).

The modulation of the substrate properties in one or two directions can create ultrapure (Q1D) or (Q0D) systems on base of SE [3]. Because both the small mass of free electron and low temperature in such systems along surface take place the quantum effects as well as classical one. The crossover here is commensurability the of mesoscopic size with electron wave length ( ~ 1000 angstroms at the low temperatures) which changes as $T^{-\frac{1}{2}}$.

The system of SE is simply organized: source of free electrons charges the structured substrate situated under helium in capacitance gap. The electrons density $n$, here is defined by exterior electric potential V, when electric layer exclude field of this potential. The density can be calculated from relation $n \cdot e = C \cdot V$ (C is value of capacitance between electron layer and lower plate). Such system is easy operated by the electric and magnetic field in broad interval of electron density. The experiments show that the Q1D SE systems as well subject of investigation as high-sensitive tool for study of surrounding condensed matter.

In this work is used Q1D - SE system which is proposed in [4] and realized in [5] where is charged the surface of liquid helium in grooves of profiled substrate. The theoretical description has been performed in [6]. The electrons are localized across channels in parabolic potential dimple $U(y) = m\omega_0^2 y^2 / 2$. Solution of the Schrödinger's equation give the wave function and the localization size of the electron in basic state are $\psi_o(y) = 1/(\pi^{1/2} y_0) \exp(-x^2 / 2y_0^2)$ and $y_0^2 = \hbar/(2\pi m \omega_0)$, accordingly (here vibration spectrum is $\omega_0^2 = eE_\perp / (mR)$ ). In experiments the value $y_0$ may achieve lower than 100 nm the magnitude which coincides with channel width.

**Experimental results and discussion.**

Here has be investigated the conductivity of narrow electron stripes profiled substrate which was negative charged. The measurements were performed by a technique which uses a capacitive couple of electrodes with electron layer [7]. The construction of cell (fig.1) and experimental steps were like the work [8].

The substrate is rows of dielectric threads in diameter 100 μm , which are placed on thin glass plate with size 10×5 mm². The substrate is situated on the two measurement electrodes. From free source are electrons emitted on substrate at temperature 1.3 K (the electrons is thermolized at these temperature) at scanning electric field $E_\perp$. Difference of levels between substrate and liquid is H, set the radius of liquid in substrate grooves $R = \sigma / \rho gh$ which typically was 35μm for our experiments. The strip width was estimated by same magnitude.

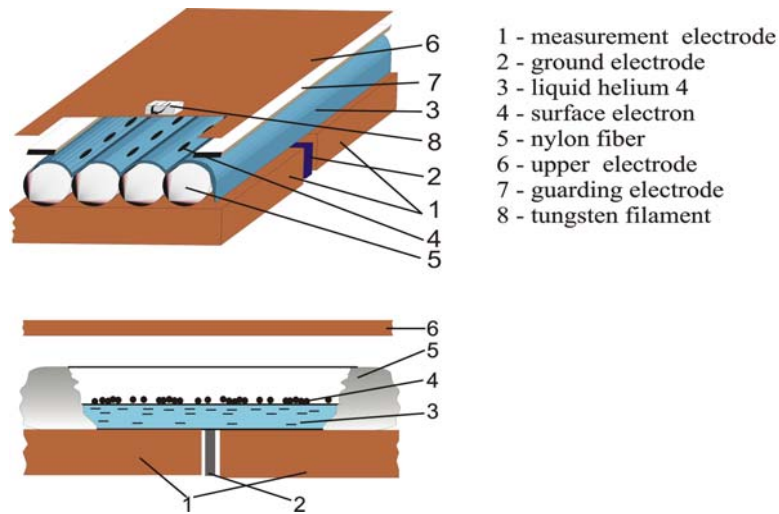

Fig.1. The experimental cell for study of Q1D SE system in negative charge presence on substrate.

In the field $E_\perp = 490 V/cm$ the linear concentration of SE was $2\cdot 10^4$ см$^{-1}$. The potential dimple for surface electron over the center of the liquid helium can be estimate as φ = e E⊥ δ (δ is the size of an arrow of a deflection, this value is defined geometrically, taking into account curvature radius in the channel and diameter of a threads). It was near $10^4$ K. Distance between interenergy in equidistant spectrum of one-dimensional electron system at these parameters is ~ 0,1 K and it is essential lower the temperatures in the experiment. At Boltzmann's energy distribution, the electrons occupied the high levels too.

The charge and its potential on a substrate could be established by electron breakdown at known concentration of SE.

Conductivity and mobility SE at known concentration were defined by look-in-amplifier on values 0-degree and 90-degree a component of a measuring signal [8]. The frequency and value signal were 20 kHz and 5-30 mV, correspondently.

On fig. 2 is shown the dependences of electron conductivities in Q1D to system from temperature.

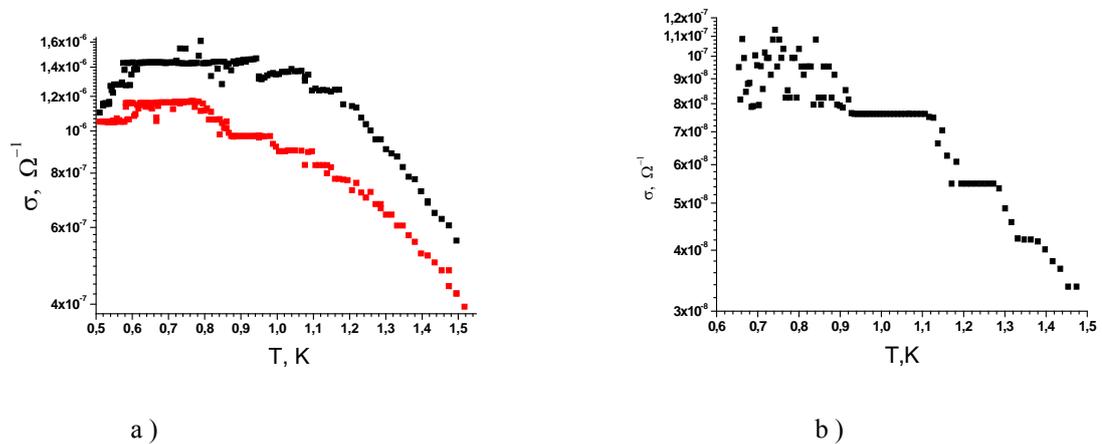

Fig. 2. Dependences of conductivity SE in Q1D to system from temperature (detail description in text).

Fig. 2a shows the conductivity SE is a function of temperature at density ~ $10^8$ cm$^{-2}$ and substrate charge potential ~30V. The first dependence on time is upper curve and bottom curve is same dependence at increasing temperature. It can see ladder-like temperature dependence of σ at T ≤ 1,3. On the bottom curve the steps are more brightly expressed. Extent of steps is more expressed at the decreasing temperature. It is remarkable that from experiment to experiment the parameters of steps are changed, but in separate experiment after a number of thermo cycles the character of steps remains almost constant. On fig. 2b similar dependence after a number of thermo cycles at SE concentration ~ $10^7$ cm$^{-2}$ is presented. The charge potential on a substrate here is near 15 V. Ladder-like picture of conductivity doesn't depended on leading potential in a range 2-150 mV. It is to notice that differential of conductivity between steps exceeds this magnitude for case of absence of a charge on a substrate. It is necessary to notice too that in case of a substrate with rather big surface inhomogeneities (the degraded substrate) step-like dependence of conductivity from *T* isn't observed.

In discussion of results we must note that electrons on profiled substrate form the lines along borders of a conducting channels . This is caused, on the one hand, by displacement of electrons down on cylinder surface of threads with potential is $\delta V = eE_\perp z$. And on the other hand the electron lines are stabilized in the displacement by the electrostatic mutual interaction with potential $\varphi = (\tau/2\pi e)\ln(1/r)$, here $\tau$ is linear charge density, $r$ is radial coordinate. The presence of SE in the conducting channels leads to some changing of this interaction.

The ladder-like temperature dependence of conductivity can be explained by presence of the fluctuation potential (FP) induced by electrons on substrate which influence on SE transport in channels. The FP is determined as $\Delta \cdot (\ln(1+R^2/r_0^2))$ with energy scale: $\Delta = e^2(\pi n_s)^{1/2}/\varepsilon$ ($n_s$ is electron density in stripes, $R$ is screening radius). At $n_s = 10^{11} - 10^{12} m^{-2}$ and at $\varepsilon \sim 5$ (for glass substrate), the value of $\Delta$ is commensurable with temperature and at lower temperature takes place a quantum transport of SE. The quantum transport here is analogical to transport in solid state systems which studied enough broadly (for example [9]).

The alternative explanation can be next. At the localization of part of surface electrons in minimum of FP at lowering temperature the conducting stripes go to narrowing. On the other hand the electron wave length is changing as $T^{-1/2}$. So at some temperature the electron wave length will commensurate with channel size. And this fact leads to quantum carry. But size steps of ladder-like conductivity don't correspond to number of the resistance quantum's $\hbar/e^2$, of quantum wire[10].

**Conclusion**

We investigated the conductivity of surface electrons in narrow stripes on helium at negatively charged substrate. At temperature lower than 1.3 K take place a step-like temperature dependence of conductivity. This feature can be explained by predominate of the quantum electron transport at definite temperatures. Last is caused by fluctuation potential which is induced by charge on profiled substrate. The alternative explanation is commensuration of electron wave length with the conducting channel width.

This work is supported partially by grant UNTC 5211.